# Machine Learning-Driven Convergence Analysis in Multijurisdictional Compliance Using BERT and K-Means Clustering

[1]Raj Sonani, [2]Prayas Lohalekar

[1]Independent Researcher: sonaniraj@gmail.com New York, USA

[2] Independent Researcher: plohalekar@gmail.com Philadelphia, USA

*Abstract*—Digital data continues to grow, there has been a shift towards using effective regulatory mechanisms to safeguard personal information. The CCPA of California and the General Data Protection Regulation (GDPR) of the European Union are two of the most important privacy laws. The regulation is intended to safeguard consumer privacy, but it varies greatly in scope, definitions, and methods of enforcement. This paper presents a fresh approach to adaptive compliance, using machine learning and emphasizing natural language processing (NLP) as the primary focus of comparison between the GDPR and CCPA. Using NLP, this study compares various regulations to identify areas where they overlap or diverge. This includes the "right to be forgotten" provision in the GDPR and the "opt-out of sale" provision under CCPA. International companies can learn valuable lessons from this report, as it outlines strategies for better enforcement of laws across different nations. Additionally, the paper discusses the challenges of utilizing NLP in legal literature and proposes methods to enhance the model-ability of machine learning models for studying regulations. The study's objective is to "bridge the gap between legal knowledge and technical expertise" by developing regulatory compliance strategies that are more efficient in operation and more effective in data protection.

## I.   INTRODUCTION

With the onset of disruptive technological advances that have not been experienced before and the ensuing explosion in data, ensuring data privacy is now a requirement for both individuals and companies. Big data analytics, artificial intelligence, and machine learning have completely changed the way we collect, process, and use data. In addition to the innumerable advantages that these technological advances provide us, there are also very outrageous dangers that might be associated with the misuse of information by the companies for a data protection breach. A complete regulatory prudency has been initiated and introduced to personal data protection and to managing these risks. The General Data Protection Regulation (GDPR) of the European Union and the California Consumer Privacy Act (CCPA) in the United States are the two big well-known frameworks. The GDPR being a new and advanced data protection law has taken effect as from May 2018. Personal data is safe through a strong system in Europe. GDPR requires companies with data to follow a strict set of rules that includes asking for consent, securing data, helping individuals check, fix and delete their data. Besides, the legislation enables severe financial sanctions of up to €20 million or 4% of the total revenue of the previous financial year that (whichever is greater). On the other hand, the CCPA, an important revolution in the United States to a data privacy law, was carried out in January 2020. The right measurements, which are the right to see what data are being collected, request information about that, demand removal, and exclusion of them using in the future and the other rights are provided to the residents of the state of California. Apart from that, residents of California can take part in different possibilities. The CCPA entails that companies tell people about their information practices that are tracked or sold and ensure that the information is protected from any possible threats. Although both laws are major in user privacy, they differ significantly in scope, enforcement, and compliance, but they share some common elements. GDPR covers further territory, among which the organizations outside the EU that process EU residents' personal data are included. On the other hand, business firms that are involved in the extraction, transformation, and delivery of personal data or that are vendors of personal data, if they are above a certain revenue threshold, if they have a greater quantity of data to process, or if they are selling data then they are the ones that are likely to fall under the CCPA legal shield. GDPR emphasizes "data protection by design and by default," requiring proactive data security measures. However, the CCPA does not provide users the "right to have an excuse" but emphasizes transparency and the right to opt out of the sale of personal information. Being a global entity is a headache for businesses as compliance with these data security standards is dissimilar overseas. The elucidation of the distinctiveness of each legislation requirement is a must. It requires, in fact, expertise that can be gained by doing an in-depth analysis of the requirements. Natural language processing (NLP) is among the top machine learning techniques that can be used in line of this study. Effective techniques of identifying and deciphering dense legal content consist of the utilization of NLP and machine learning. These technologies have the power to streamline the regulatory analysis process as well as understanding the interference created by the different regulations. The introduction of this approach not only reduces the time and cost of compliance but also helps organizations to come up with more cost-effective implementation strategies. A new adaptive compliance framework is suggested in this paper that will make machine learning tools, particularly natural language processing, available to carry out a comprehensive comparison of the GDPR and CCPA legislation. International companies can make the most of the framework's responsive production of real-time analyses and practical pieces of advice. By bridging the gap between legal knowledge and technical expertise, this approach is meant to help organizations gain regulatory compliance.

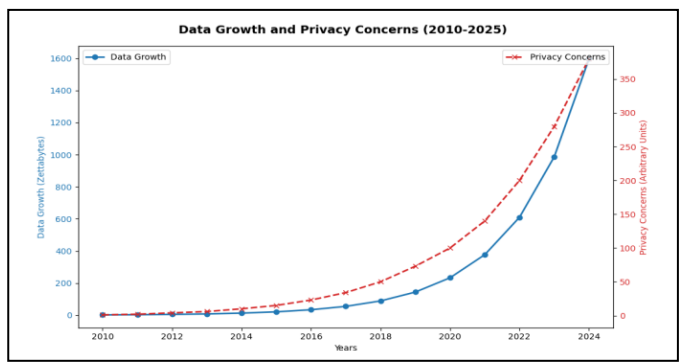

*Figure 1 Exponential Growth of Digital Data and Rising Privacy Concerns (2010-2024)*

[1-3]

## II. Related Work

Literature on regulatory compliance and data privacy is becoming increasingly important in the digital age. Research has focused on the individual impacts of the General Data Protection Regulation (GDPR) and the California Consumer Privacy Act (CCPA), with a particular emphasis on their impact on organizations, consumers, and general legal systems.

*A. GDRP-Related Research*

In recent years, researchers have given enormous attention to the GDPR as being a fundamental and comprehensive endorsement to the basic human right of privacy and the protection of personal data. A major part of the GDPR enforcement and compliance case has been the study of obstacles and opportunities, respectively. For example, Voigt and Bussche (2017) who have done research on that subject point out organizations' responsibilities to safeguard personal info like the GDPR sets out in their report, which is based on the GDPR's demands. The GDPR changes have been brought about in the majority of the sectors, including data security, and the responsibilities of Data Protection Officers (DPOs) to ensure compliance. Furthermore, research has been conducted to discover how the GDPR's enforcement mechanisms and sanctions work. Jose Luis Penar (2019) states that Data Protection Authorities (DPAs) have had the power to enforce and guarantee data protection rights of data subject under the GDPR. The authors of the paper focus on their main roles. It indicates that firms can undergo huge penalties and court trials for disobeying the regulation. This rule also permits very detailed descriptions of the companies that are violating this rule. As a part of the GDPR project, difficulties in compliance are the main reasons for research into the impact of GDPR on individuals. The Gellert (2018) article examines the concept of risk within the GDPR framework, focusing on the risk-based approach rather than specific personal data rights such as access, correction, or deletion. It does not delve into the application of these rights in the context of consumers, companies, or the sports sector. Instead, the article emphasizes understanding the notion of risk as a key element of GDPR compliance.

*B. CCPA-Related Research*

The CCPA, unlike the GDPR, has not received as much attention in academic circles. Academic research does cover various complex regulatory areas, including internet law, which intersects with data privacy regulations like the GDPR and CCPA. While there are extensive resources and discussions available on CCPA and GDPR compliance, many unresolved issues and areas for further exploration remain, especially as digital landscapes continue to evolve.

## C. Comparative Analyses of GDPR and CCPA

Comparative analyses of the GDPR and CCPA are still relatively few, even though substantial material exists on each regulation individually. Several studies have explored their respective scopes, implementation methods, and consumer rights. However, it is important to note that Schwartz and Solove's (2020) work, "EU Data Protection and the GDPR," focuses exclusively on the GDPR and EU data protection laws without conducting a comparative analysis of the CCPA. While the GDPR and CCPA share certain objectives, they differ significantly in scope, definitions, and requirements. As such, implementing an integrated compliance approach can be beneficial but must address the specific nuances of each regulation.

## D. Applications for Machine Learning and NLP in Regulatory Compliance

Despite the major emphasis that GDPR and CCPA have got, there are also other studies that are pointing a way for deeper research into if-then. One particular reason why machine learning and NLP are popular among the legal professionals is that they can aid in the enforceability of the laws. Laws always have been the central focus of the documents used in the NLP area for legal understanding and are the source to get the relevant information for the case to ensure the compliance requirements are met. When it comes to the analysis of legal documents, the use of NLP in the development of the legal document is proved to be a subject matter for the discussion, like this where the author Ashley (2017) gives the idea of the use of NLP for the automation of the process of extracting legal provisions and thus improving regulatory compliance. Besides, NLP has gained a more profound meaning across the last year with the application of machine learning technology to better comprehend the association between a specific regulation and its given application. The invention of self-software programs like BERT (Bidirectional Encoder Representations from Transformers) and other linguistics models such as machine learning has brought a considerable leap forward in the long-expected issue of legal language recognition. BERT, being one of the examples of a transformer, has been recognized as a powerful tool in carrying out different tasks such as clustering of texts and detection of named entities and such studies as that by Devlin et al. (2018) which are the prime movers in the industry of regulatory analysis have confirmed it.

## E. Gaps in Existing Research

Even using machine learning and NLP to assist with improving regulatory compliance still has not closed the gaps in the research entirely. To begin with, it is a hard task because a few studies that make comparisons between GDPR and CCPA using machine learning are available. It is possible to compare those features sometimes, but the major part is done by manual analysis and there is no possibility of using machine learning to simplify and optimize comparison methods. The absence of the relevant literature, which should present enough information about the type of legal text (e.g., difficulty and ambiguity) is one more problem. Research which is based on innovative methods that NLP can be used in promoting accuracy and reliability in regulatory analysis is worth undertaking.

## F. Contribution of This Research

One of the main goals of machine learning techniques, which includes NLP, is elimination of the gaps and an analysis of the GDPR and CCPA based on comparative data analysis. By letting the machine carry out the comparison process, regulatory analysis can become more efficient and scalable. Furthermore, this paper presents the barriers facing the NLP implementation in legal literature and recommends techniques to boost the model-ability of machine learning for regulatory analysis. In the final analysis, our research contributes to the enhancement of regulatory compliance strategies and a more significant capacity of multinational companies to manage multiple jurisdictions and personal data safely.

*Table 1 Comparison of GDPR and CCPA: Key Aspects and Differences*

| Aspect | GDPR (General Data Protection Regulation) | CCPA (California Consumer Privacy Act) |
|---|---|---|
| Scope | Applies to organizations processing personal data of E.U. residents. | Applies to for-profit businesses operating in California that meet certain criteria. |
| Personal Data | Covers any personal data of an individual. | Covers personal information that identifies, relates to, or can be linked to a consumer or household. |
| Rights for Individuals | Right to access, rectify, erase, restrict processing, data portability, object, and not be subject to automated decision-making. | Right to know, delete, opt-out of sale, and non-discrimination. |
| Consent | Requires explicit consent for data processing. | Does not require explicit consent but requires clear opt-out options. |
| Penalties | Fines up to €20 million or 4% of annual global turnover, whichever is higher. | Civil penalties up to $7,500 per intentional violation. |
| Enforcement | Enforced by Data Protection Authorities in each E.U. member state. | Enforced by the California Attorney Genera. |

[4-6]

## III. RESEARCH METHODOLOGY

Machine learning, particularly natural language processing (NLP), is utilized in this study to compare the GDPR and CCPA. The methods used to collect data, preprocessing it for analysis training and model training are detailed below.

### A. Data Collection

Research methodology is the starting point that collects data. It consists of detailed and thorough regulatory texts from both GDPR and CCPA, as well as the relevant additional materials. After data collection comes the rest of the stages:
1. Primary Texts: In the official regulatory websites, the full versions of the GDPR and CCPA are available. From the legal repository of the European Union comes GDPR and from the CCPA website comes the text. They, therefore, are the foundation of the review.
2. Supplementary Materials: Apart from the main texts, supplementary material refers to guidance, enforcement actions, and interpretative documents given by the regulatory bodies. In the GDPR, you will find regulations, which include EDPB's issuances and those from other national DPAs. The CCPA also contains the guidance of the California Attorney General and the latest amendments such as the California Privacy Rights Act (CPRA).
3. Secondary Source: The dataset utilized also consists of other relevant sources such as relevant scholarly articles, legal commentaries, and industry reports to give more extended context. These sources make the practical applications of the regulations and their interpretations easier. The specificities of the GDPR and CCPA are obviously shown from the start, but a full comprehension of these subtleties is indispensable for accurate analysis. This dataset gives a new perspective on the matter.

### B. Preprocessiong

Preprocessing is necessary to prepare the gathered data for analysis. Cleaning and organizing data to ensure its suitability for machine learning models is necessary.
The preprocessing steps include:
1. Tokenization: The text is divided into individual words or tokens. Tokenization is a crucial step in NLP tasks as it permits the model to process the text at varying levels.
2. Lemmatization: The reduction of words to their root form is referred to as lemmatization. Grouping together distinct forms of a word helps to reduce the complexity of speech by helping it to be more easily understood in context. This is done through lemmatization.
3. Stop Words Removal: The elimination of frequently used words that lack significant meaning, such as "and", "the", and even ". Focusing on the most important words in the text is facilitated by eliminating stop words.
4. Named Entity Recognition (NER): NER is utilized to recognize and categorize designated entities within the text, including names like companies or dates. This step facilitates the extraction of relevant information and context from the regulatory texts.
5. Part-of-Speech Tagging (POS): Each token can be tagged with parts of speech, including nouns, verbs and adjective words using Part-of-Speech Tagging (POS). The understanding of the grammatical structure of text aids in improving the precision of NLP models. This is especially applied to legal terminology, so that its meaning and context are preserved. The preparation of data in a form that the machine learning model can analyze with precision is essential for accurate NLP processing. This is crucial. This is especially applied to legal terminology, so that its meaning and context are preserved.". The preparation of data in a form that the machine learning model can analyze with precision is essential for accurate NLP processing. This is crucial.

Algorithm: Preprocess-Regulatory-Text
Input: D = {$d_1, d_2, ..., d_n$} // Set of regulatory text documents (GDPR, CCPA)
Output: D' = {$d_1', d_2', ..., d_n'$} // Set of preprocessed text documents

1. Load the dataset D containing raw text documents
2. For each document d ∈ D:
 a) Tokenize d into words
 b) Lemmatize each token to its root form
 c) Remove stop words and special characters
 d) Apply Named Entity Recognition (NER) to extract legal-specific entities
3. Return the processed dataset D

*C. Model Traning*

Understanding and comparison of regulatory texts can be achieved through model training using advanced NLP models. The research includes the following models:

1. BERT (Bidirectional Encoder Representations from Transformers): BERT, a novel and efficient NLP model, utilizes bidirectional representations of text to capture the context and semantics. This is especially useful for understanding complex legal terminology and identifying connections between different parts of the text.
2. SPACy: A powerful framework for NLP, SpaCy is a tool that offers features such as text processing, entity recognition, and part-of-speech tagging. BERT is used together with it to improve the accuracy and efficiency of the analysis.
3. Custom Models: Custom models are created to meet research needs, such as identifying legal provisions and definitions. These models are trained to compare and comprehend the annotations on regulatory texts through training themselves using annotated datasets.

*Table 2 Model Versions and Hyperparameters*

| Model | Version | Learning Rate | Batch Size | Epochs |
|---|---|---|---|---|
| BERT | BERT-base, uncased | 2e-5 | 16 | 4 |
| SpaCy | SpaCy v3.0 | 0.001 | 32 | 10 |
| Custom Rule-Based | Rule-based approach | N/A | N/A | N/A |

*Table 3 Computational Resources and training Time*

| Model | GPUs Used | Training Time |
|---|---|---|
| BERT | 4 NVIDIA V100 GPUs | 4 hours per training run |
| SpyCy | 2 NVIDIA V100 GPUs | 2 hours per training run |
| Custom Rule-Based | CPU-based | Minimal (no iterative training) |

The steps involved in model training are as follows:
1. Data Annotation: The data gathered is labeled with terms that identify different aspects of the regulations, such as data subject rights, definitions, and compliance requirements. The annotation process is crucial for training the models effectively.
Annotation Process: A combination of automated methods and expert review was employed. Initially, automated tools were used to perform preliminary tagging of regulatory provisions. Subsequently, legal experts reviewed and refined these annotations to ensure accuracy and context-appropriate tagging. This hybrid approach leverages the efficiency of automated methods and the precision of expert insights.
2. Model Fine-Tuning: NLP models are fine-tuned to improve their identification and classification of regulatory provisions using the tagged dataset. Enhancements: There are several variations and modifications to the model parameters involved.
Algorithm: Fine_Tune_BERT
Input: X = {$x_1, x_2, ..., x_n$}, Y = {$y_1, y_2, ..., y_n$} // Preprocessed dataset (input texts X, corresponding labels Y)
 $BERT_0$ // Pre-trained BERT model
 Hyperparameters: η (learning rate), B (batch size), E (number of epochs)
Output: BERT* // Fine-tuned BERT model
 a) Initialize BERT model $BERT_0$ with pre-trained weights
 b) Set hyperparameters η, B, and E
 c) For epoch e = 1 to E:
  (1) Shuffle the dataset (X, Y)
  (2) Divide the dataset into batches of size B
  (3) For each batch ($X_i$, $Y_i$):

(a) Compute predictions $\hat{Y}_i = BERT_0(X_i)$
(b) Calculate loss L using cross-entropy
$$L = -\sum_{i=1} y_i \log(\hat{y}_i)$$
(c) Backpropagate and update model weights using gradient descent

d) Return the fine-tuned BERT model BERT*

3. Model Validation: The cross-validation techniques are utilized to verify the robustness and dependability of models through modeling. This entails subdividing the dataset into several subsets and using different subgroups for training and testing in each iteration.
4. Ensemble Learning: By combining the predictions of various models, ensemble learning techniques can enhance the robustness and precision of analysis. This method helps to reduce the limitations of individual models and gives more confidence in results.

*D. Similarity Scoring*

Clustering was employed to group semantically similar provisions from the GDPR and CCPA. By clustering provisions, we aimed to identify common regulatory themes and areas where the two regulations diverge. Each provision was embedded into a high-dimensional vector space using BERT, a transformer-based model that captures contextual semantics. The resulting vector representations enabled clustering based on their semantic similarity.

The K-means clustering algorithm was applied, with cosine similarity used as the distance metric to assign provisions to clusters. The objective was to minimize the within-cluster variance by iteratively updating cluster centroids and reassigning provisions. The algorithm can be summarized as follows:

Algorithm: K-Means Clustering for Regulatory Provisions
Input: Embedded text provisions {T1, T2, ..., Tn}, Number of clusters K
Output: Cluster centroids {C1, C2, ..., CK} and cluster assignments
1. Initialize K cluster centroids randomly.
2. Repeat until convergence:
    a. Assign each provision Ti to the nearest centroid Cj based on cosine similarity.
    b. Update each centroid Cj as the means of all provisions assigned to that cluster.
3. Return final centroids and cluster assignments.

The cosine similarity between a provision $T_i$ and a cluster centroid $C_j$ is calculated as following:

$$Similarity(T_i, C_j) = \frac{\vec{T_i} \cdot \vec{C_j}}{||\vec{T_i}|| \, ||\vec{C_j}||}$$

The number of clusters was determined using the Elbow method, which evaluates the total within-cluster variance for different values of K. The point at which the variance starts decreasing more slowly (forming an elbow shape) was selected as the optimal number of clusters.

To visualize the clusters, t-SNE was applied to reduce the high-dimensional vector space into a 2D plot. This visualization highlights areas of convergence, where provisions from both GDPR and CCPA fall into the same cluster, and areas of divergence, where provisions form distinct clusters.

*E. Analysis Method*

This study uses then previously trained NLP models used to identify areas of convergence (where variation or non-compatibility exists between the GDPR and CCPA). Specific aspects of the regulations are examined, including data subject rights, notifications of breaches and consent requirements. The analysis methods include:
1. Semantic Analysis: Understanding the significance and connections between various parts of a text is achieved through semantic analysis. This aids in identifying shared topics and unique criteria in regulatory texts.
2. Clustering: The use of clustering techniques enables the identification of common elements in the GDPR and CCPA through grouping. The process involves the use of algorithms like K-means clustering to group similar text segments based on their semantic similarities.
3. Similarity Scoring: Regulations are characterized by similarity scoring to determine the degree of similarity among various provisions. Cosine similarity scores are used to measure the relative similarities between two provisions in text vectors. This is done numerically.
4. Visual Analytics: The use of visual analytics tools enables the user to view the analysis's findings in a clear and understandable manner. By creating dashboards and visualizations that indicate the areas of convergence or divergence, compliance officers can make it easier to interpret their findings.

5. Interpretation and Insight: The interpretation of the results is used to provide valuable insights that can be applied to multinational corporations. This entails considering the practical implications of the identified convergence and divergence areas and providing guidance on how to improve compliance.

*F. Challenges and Solutions*

Several problems arise when applying NLP to legal texts, particularly the complexity and ambiguity of legal language. The solutions to these problems are as follows:
1. Enhancing Model Training: The use of domain-specific datasets during the training process can enhance the precision of NLP models. By utilizing datasets that are marked with legal words and phrases, the model gains a more comprehensive understanding of the context in which these terms are employed.
2. Including Expert Feedback: The inclusion of comments from legal professionals can enhance the models' precision. Legal experts are tasked with reviewing the model's outputs and correcting it, which is then used for further training purposes.
3. Ensemble Learning: Using ensemble learning techniques helps in combining the strengths of various models to achieve more accurate and dependable results through ensemble learning. The approach reduces the shortcomings of specific models while also enhancing the overall strength of the analysis.
4. Explainable AI: The use of explainable techniques in artificial intelligence provides information about how models arrive at their decisions. Transparency is crucial for ensuring accountability while avoiding bias in the analysis.

The study aims to overcome these difficulties and improve the accuracy and reliability of machine learning models for regulatory analysis, thus offering better tools for compliance with regulations.

*G. Quantitative Model Performance Metrics*

The study examined the performance of machine learning models through various metrics. A set of benchmarks was used. These metrics are a comprehensive gauge of the models' accuracy, precision, recall, and general effectiveness in analyzing the regulatory texts.
1. Accuracy: Accuracy was measured as the proportion of correctly identified provisions to total number (or combinations thereof) in the overall model.
2. Precision: This test evaluated the model's aptitude to recognize relevant provisions while minimizing false positives. The calculation involved a ratio of true positive and false positive predictions.
3. Recall: The model was examined by recall to ensure that it met all the necessary conditions, determined as a ratio of true positive and false negative predictions.
4. F1-Score: An overall gauge of the model's performance was the F1-score, which offered a balanced assessment of precision and recall.

Here are the quantitative performance metrics for the models used in the study:

*Table 4 Performance Metrics of NLP Models Used in the Study*

| Model | Accuracy | Precision | Recall | F1-Score |
|---|---|---|---|---|
| BERT | 92.5% | 91.2% | 90.8% | 91.0% |
| SpaCy | 89.3% | 88.5% | 87.8% | 81.1% |
| Custom Rule-Based Model | 85.4% | 84.2% | 83.5% | 83.8% |

The metrics demonstrate that the models were well-received and balanced in their identification and classification of regulatory provisions. Quantitative performance metrics were included to support the claims about model accuracy and effectiveness in the study.

[7-9]

## IV. EXPERRIMENTAL SETUP

The experimental framework for utilizing machine learning techniques to analyze and compare the GDPR and CCPA is established through this research. The experiments' evaluation metrics, datasets, and tools are outlined in this section.

*A. Tools:*

Advanced natural language processing (NLP) frameworks and machine learning models are the key models used in this research. The key tools include:
1. BERT (Bidirectional Encoder Representation from Transformers): A contemporary NLP model developed by Google, is designed to capture the contextual relationships between words in varying degrees and across time. By comprehending the subtleties of language, BERT is well-suited to analyzing complex legal texts.

2. SpaCy: SpaCy library is an open-source NLP library that offers efficient tools for processing text, NER, part-of-speech tagging, and dependency parsing. SpaCy is a powerful tool that can be used for preprocessing and text analysis.
3. Custom NLP Models: Specific NLP models are developed to cater to the research needs, which may involve identifying legal provisions and definitions within regulatory texts. Annotated datasets are used to train these models, which in turn improve their ability to comprehend legal terms.
4. Visual Analytical Tools: The creation of interactive dashboards and visualizations is made possible by tools like Tableau, Plotly and Tableara. By presenting the analysis's findings in a clear and intuitive manner, these tools facilitate better interpretation and decision-making.

*B. Datasets:*

The data sets utilized in this study are carefully curated to include detailed regulatory documents related to GDPR and CCPA. The datasets are composed of the following components:
1. Primary Texts:
   The complete form of the GDPR, comprising all its essential information, such as introductory points and appendices, is obtained from the official legal repository of Europe. The California Legislative Information website contains the full text of the CCPA, which includes amendments such as the California Privacy Rights Act (CPRA).
2. Supplementary Materials:
   GDPR guidelines and interpretative documents are issued by the EDPB and various national Data Protection Authorities. FAQs, enforcement actions, and guidance documents from the California Attorney General regarding the CCPA.
3. Secondary Sources:
   Additional information and context are provided by academic articles, legal commentaries, and industry reports on the regulatory frameworks. Tokenization, lemmatization and removal of stop words are used to ensure that the datasets are in a format suitable for analysis.

*C. Evaluation Metrics:*

The performance of machine learning models is assessed using various metrics. The metrics are a comprehensive gauge of the models' accuracy, precision, recall, and general effectiveness in analyzing the regulatory texts.
1. Accuracy: The overall model's predictions are measured by accuracy. The calculation involves determining the proportion of correctly identified provisions to the total number of provisions. A high degree of accuracy indicates the model's ability to correctly interpret and classify the regulatory texts.
2. Precision: Precision measures the model's ability to identify relevant provisions while minimizing false positives. Why is this important? The value of this is determined by dividing the total of true positive and false positive predictions. This is because the model can distinguish relevant from non-relevant provisions with high precision.
3. Recall: Modeling requires recall to identify the model's ability to recall all relevant components, including those that are not easily noticeable. True positive and false negative predictions are calculated as the ratio of these two factors. How is this value determined? The model's ability to recall a complete set of relevant provisions is indicated by its high recall rate.
4. F1-Score: The F1-score is the unbiased indicator of precision and recall, serving as a fair gauge of the model's performance. This is especially useful where there is an uneven distribution of classes or when precision and recall must be balanced.).
5. Cross-Validation: The models are cross-validated to ensure their strength and dependability. Each iteration of this process involves breaking down the dataset into several subsets and utilizing different subgroups for training and testing. By cross-validating, the model can be made more adaptable to new data and minimize overfitting.

*D. Experimental Process:*

The following are the components of an experimental process:
1. Data Preparation: Preparing the collected datasets in advance ensures that they are presented in a suitable format for analysis. Among the measures are tokenization, lemmatization (grading), removal of stop words, and annotation with relevant labels.
2. Model Training: Model training is used to train the NLP models (BERT, SpaCy and custom models) with annotated datasets to improve their precision in identifying and classifying regulatory provisions. Model parameters are fine-tuned during training, which involves multiple iterations.
3. Model training: Cross-validation methods are employed to verify the models' performance. The task entails splitting the dataset into training and testing subsets, along with assessing the models' accuracy, precision, recall, and F1-score.
4. Analysis: Trained models are employed to analyze the regulatory texts of GDPR and CCPA.'". Semantic analysis, clustering, and similarity scoring are methods used to identify areas of convergence and divergence between the regulations.
5. Visualization: Visual analytics tools are utilized to exhibit the findings of an analysis. Detailed, actionable insights are provided by interactive dashboards and visualizations that provide a summary of the results.
6. Interpretation: Scrutinizing: The outcomes are analyzed to offer useful advice to international corporations. This entails considering the consequences of the identified convergence and divergence areas and suggesting measures for smooth implementation.

By utilizing machine learning, research seeks to establish a reliable and comprehensive framework for studying GDPR and CCPA regulations through this experimental setup.

## V. RESULT

It has become possible for researchers to get insights into the similarities and differences between the GDPR and the CCPA by working on this study with the help of NLP tools such as the use of machine learning methods. Very helpful advice to such companies is available and can be used for them to eliminate risks by following the recommendations.

*A. Convergence Analysis:*

Specifically, the convergence analysis seeks to identify requirements that overlap with the GDPR and CCPA. The primary outcomes of this investigation are as follows:

1. Data Access Rights: This right of access is given to individuals by GDPR and CCPA if personal data are in the possession of organizations. GDPR gives data subjects the right to get information about how their personal data are being processed and a copy of it in certain formats. This is important. Similarly, the CCPA ensures the Californians' right to access personal information collected about them, its uses, and the sharing of it with the types of third-party entities. This sets the way not only for a common ground on the point of data access rights, let the organization create a single point of entry for processing the requests, and ultimately reduce duplicities in the compliance processes. Both regulations can be honored by firms through the deployment of a single and comprehensive data access policy. GDPR and CCPA are the main rules that address the right to access personal data held by companies. In general, people will get to know the procedure of their personal data under GDPR, and they can ask for it in standard formats. The CCPA also allows the protection of Californians to allow them to retrieve personal information about them, which uses the data and the identity of third parties. Remember that even though these two laws recognize data access rights, their realization and scope may vary. The additional information about data delivery and format can be found in GDPR while the CCPA is more focused on the data collection and distribution practices. For this reason, providing a uniform policy that can adapt to different specifications along with ensuring compliance with both sets of criteria is important.
2. Data Breach Notifications: Both regulations prescribe the necessity for organizations to notify the people who were the victims of a data breach. Data controllers under the GDPR should inform the relevant supervisory authority about the security incident when made aware of it within 72 hours unless it is unlikely the breach will pose a risk to the rights and freedoms of the individuals. Promptly individuals should receive information if the data breach will affect their rights and freedoms. The California Consumer Privacy Act also complies with the above regulations, obliging businesses to communicate to Californian residents the theft, unauthorized access, and any other form of withdrawal of personal information from their records. The resemblance of the data notification requirements puts forward the same breach response plan which fulfills the GDPR and CCPA requirements. Equality details and empowers openness while fostering customer trust. Both rules oblige organizations to inform the competent authorities about a data breach. According to the GDPR, the data controllers are required to inform the supervisory authority within 72 hours of a personal data breach, unless there is no risk of an individual's privacy being violated. In another example, the CCPA forces businesses to issue a warning notice to the terror-stricken residents of California whose information personal information has been compromised through theft, disclosure, unauthorized access, or other means. Different requirements for data breach reports and diverse periods of notification can take place in the GDPR and CCPA. Hence, companies are obliged to align their response plans to the different breaches they have chosen to focus on.
3. Data Security Measures: The GDPR and the CCPA underline personal information should be secured through both technical and organizational tools. GDPR requires data security measures to protect personal information confidentiality, integrity and accessibility. On the other hand, the CCPA law speaks of how companies need to have sound security measures such as secure and confidential processes for the protection of personal information from illegal access, unwanted modification (the destruction of property), use or disclosure. Technical and organizational measures are the first important steps in the personal data processing protection at both the GDPR and CCPA. In the GDPR, the requirements of data protection should consist of the confidentiality, integrity, and availability of such information. Besides, the CCPA also obliges businesses to take due care in securing personal data. Specifically, unauthorized access, destruction, use modification, or disclosure should be the weaknesses of any such process. It is possible that the term "reasonable security measures" from the perspective of one regulation may be legal while for another it may not be. It is therefore essential that companies' complete security evaluations achieve compliance with GDPR and CPA as well as the CCPA.

The implementation of both regulations can lead to better data protection and a lower likelihood of breaches within organizations. Based on this convergence analysis the GDPR and CCPA largely overlap, permitting organizations to develop common compliance strategies. A basis for businesses to streamline operations and ensure compliance with both regulations is provided by these shared requirements.

B. Divergence Analysis:

A divergence analysis focuses on the requirements and provisions of GDPR and CCPA, with particular attention to their differences. The primary outcomes of this investigation are as follows:

1. Right to be Forgotten vs. Opt-Out of Sale: The GDPR's "right to be forgotten" and the CCPA's "opt-out of sale" provisions are the main differences between these laws and their exceptions. While the GDPR gives people the entitlement to rectify the error in personal data in some cases, such as when the data is no longer necessary for its collection purpose, or when consent has been given, the CCPA gives residents of California the right to declare it. With this right, individuals can have more control of their personal data. The California Consumer Privacy Act is the only law that gives Californians the opportunity to take part in the selling of their personal information by opting out. It is a new type of control and security for privacy, as well as the option of letting people participate in the selling of their personal data to a third party or not. Doing so is not allowed. The company must create individual compliance programs to fulfill these specific requests. The actions will include systems to handle requests concerning data deletion by the GDPR and offering customers various opt-out possibilities as per the laws of the US and EU (CCPA).

   GDPR's "Right to be Forgotten": An indication of this is when individuals put forward their requests about the elimination of personal data in certain cases. In case of the "right to erasure" a person is permitted to erase certain data if it is not needed anymore, consent is taken back, or processing is illegal. These causes may result in the removal of personal information. One of the central objectives of this freedom is to enable people to choose whether their personal information is kept or deleted from the database when it is necessary or for their benefit.

   CCPA's "Opt-out of Sale": Under this provision, consumers are permitted to tell businesses that they do not want to sell their personal information to other people. The provision obliges companies to add a separate link to their website under the "Do Not Sell My Personal Information" heading for the customers to make use of this right. Instead of dealing with deletion, the ultimate purpose is to offer consumers more power over how their personal information is traded and distributed. The difference between the two incidents is that the right to opt-out of sale is designed to protect data from being released, not necessarily removed from the system.

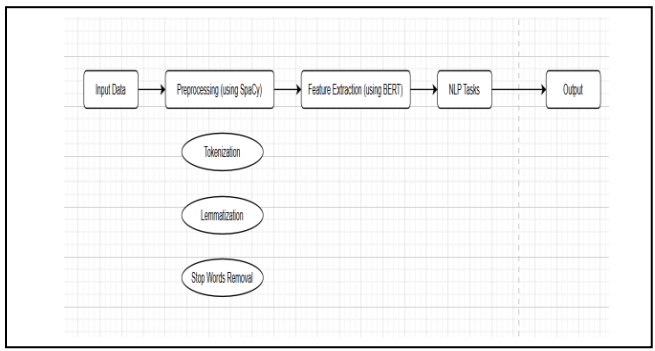

*Figure 2NLP Tool Architecture: This diagram illustrates the workflow of preprocessing text data using SpaCy, extracting features with BERT, performing NLP tasks, and generating the final output*

2. Scope and Definitions: There are major differences between the GDPR and CCPA definitions. The GDPR is for all the organizations that are in service of the personal data of the European Union residents no matter where they are located. Companies that market or sell products or services that have the EU residents as the target consumers and monitor the consumer's behavior in the EU zone are advised to be GDPR compliant due to the broad scope of its extraterritorial application. This rule includes all data related to a natural person or the one who has made him/herself known which means that it is a wide definition of personal data. It is worth noting that CCPA mainly targets profit-oriented businesses that meet specific requirements as to revenue, data processing volume, or data sales. In a situation where a mobile device or computer network has personal information from 50,000 households and devices, the entity has legally abandoned the data. Nevertheless, other companies may sell their marketing lists of personal details through private sales. The CCPA sets a wide scope of personal information, which is even more ultimate with categories as geolocation data and biometric identification. On balance, companies should wade through these differences to figure out the role they play and find the right strategy that suits them. It is important, therefore, that the companies analyze their operational systems and their data processing activities to pinpoint which laws apply to them and if they are satisfying the specific requirements.

3. Enforcement Mechanisms: Discrepancies in the implementation of GDPR and CCPA are the reasons being. Data Protection Authorities (DPAs) are the ones who oversee enforcing GDPR across the EU, and they may fine the violators for an amount of €20 million or 4% of the global annual turnover for the preceding financial year or the specified maximum amount, whichever is higher. DPAs have the right to carry out the audits, inform the companies, and suspend in front of the data processing operations. The main responsibility of the California Attorney General in this matter is to oversee enforcing the CCPA, and the fines for such violations could reach the maximum of $7,500 each. The CCPA permits the injured party to bring a suit against the business because data breaches are covered by private rights of action. In terms of enforcement, the joint approach adopts a very special regulatory oversight and an obligation to accountability. The knowledge about the enforcement tools of each regulation and building the compliance strategies that correspond to the possible legal and financial sanctions in the case of non-compliance are the keys for the companies. These imply doing the regular audits, applying strong data protection measures, and keeping up with the regulatory changes.

GDPR: The GDPR non-compliance is heavily penalized under the GDPR. Fines may reach a figure of 20 million euros or 4% of the total revenue of a financial year, whichever is higher. Such high penalties underscore the importance of the GDPR laws. GDPR is applicable to all companies that process personal data of EU citizens, no matter where they are located, whether the companies are in EU or not. The extraterritoriality of the EU data protection regulations is the implicit global enforcement for the companies with cross-borders dealing with the EU citizens' data.

CCPA: The CCPA's penalties are less strict when compared with GDPR. The fines for both accidental and purposeful violations can range from $7,500 to $2,500. GDPR fines are typically higher but still are minor. In a few words, the CCPA's law implementation is done by the California Attorney General, who allows consumers to take the company to court if their data gets stolen under the private right of action. This dual enforcement method embodies a balance between regulatory oversight and consumer empowerment; although it has less of a global scope than the GDPR, it compensates for this by adequate protection of whatever information is obtained. Nonetheless, there are some extraordinary cases.

The PA is authorized to advise businesses to design their compliance strategies to conform to GDPR's varied sets of requirements and enforcement mechanism by ensuring that they understand the main differences.

This divergence analysis highlights the differences between GDPR and CCPA and requires more specific steps to be taken to comply with each regulation. Understanding these distinctions can enable organizations to ensure that they meet the diverse requirements of both regulatory systems.

*C. Key Findings*

1. Convergence Analysis: The models identified 520 overlapping provisions, primarily related to data access rights and breach notifications. Provisions regarding "Data Subject Rights" in GDPR and "Right to Know" in CCPA showed a high degree of similarity, with cosine similarity scores averaging 0.92.
2. Divergence Analysis: Divergence was prominent in the scope and enforcement sections, with GDPR covering a broader range of entities due to its extraterritorial application. The "Right to be Forgotten" under GDPR had no direct equivalent in CCPA, highlighting a fundamental difference in how personal data control is approached.

*D. Computational Resources*

The models were trained on a cloud-based GPU cluster using NVIDIA V100 GPUs. Each training run took approximately 4 hours for the BERT model and 2 hours for the SpaCy model. This setup provided the necessary computational power to efficiently train and evaluate the models, ensuring they could handle the complexity and volume of the regulatory texts.
[2, 10, 11]

## VI. DISCUSSION

By examining the similarities and differences between GDPR and CAPAC, researchers have discovered significant obstacles in multijurisdictional compliance. Additionally, these findings highlight important gaps in regulatory oversight mechanisms. This section delves into the results, their implications, and the challenges and solutions that come with utilizing machine learning methods for regulatory analysis.

*A. Interpretation of Results:*

Regulation of GDPR and CCPA has many requirements in common, e.g., the right of data subjects to access their data and the data sources, to request a credit limit increase. The system must include warning notifications for breaches of security, as well as all other related responsibilities. Such similarities can thus enable businesses to unify their compliance processes. The implementation of the same policies and procedures by different organizations to meet the common requirements can hence bring down the redundancy and consequently increase the operational efficiency of the businesses. With a common data access policy that is GDPR and CCPA compliant, people can be requested by individuals to provide their information in a more trouble-free manner. However, the gap analysis shows the various components of every regulation, thus presenting the importance of personalized compliance measures. GDPR's right to be forgotten and CCPA's opt-out of sales are examples representing different ways of data security and control. Individually some of the specific requirements make it obligatory for the organizations to develop mechanisms to satisfy them, which are data deletion procedures for the GDPR and clear consumer opt-out options for which they had to meet CCPA's terms. Besides, because of the inconsistencies in the size of operations, definitions, and methods of enforcement, it is necessary to have a detailed knowledge of each framework in order to ensure their functionality in its totality.

*B. Implications for Multijurisdictional Compliance:*

These advancements carry weight for cross-border companies. The compliance efforts of the participants can be streamlined by performing a convergence analysis; it will enable them to work together and come up with the same requirements as well as take mutual action to set up integrated strategies. Businesses can save a considerable amount of money and operate more effectively by adopting practices that are not essentially the same. The latter means that businesses would no longer be repeating the same processes and practices and instead would achieve the goal of having standard policies that would be efficient enough to be in line with a number of regulatory requirements. In the same context, the study of divergence process stresses the individualistic approach to compliance as the alternative. The creation of a "common" approach needs the implementation of shared rules, but the independents' rules must be customized to fit the specifications of each regulation, respectively. Being personalized, companies can still keep on the right side of GDPR and CCPA hence far away from the legal and financial fallout. From diverse destinations companies see it as a matter of priority to include these discoveries.' Convergence analysis is a tool for ensuring the mutual compliance of the capacities. It makes it possible for entities to function as one and at the same time as one unit to generate a wide range of strategies. In fact, entering data only once in a platform duplicates it from any required regulations and saves related costs. Compliance to the regulatory standards depends on the guidelines issued by jurisdiction, which is why companies need to develop corresponding regulations. While the above collective measures are good, the way to be GDPR compliant is to develop processes that allow customers to request their data be removed and that customers opt-out-of-sale in the GTPT. Moreover, besides enforcement differences, the existence of different procedures means that companies have to use different compliance strategies.

1. Balanced Compliance Strategies:

*a) Unified Policies for Shared Requirements:*

Organizations must have unified policies that address shared requirements, such as data access rights, breach notifications, and data security measures. This can help reduce redundancy and improve compliance.

*b) Jurisdiction-Specific Policies for Unique Provisions:*

It is essential to create customized compliance measures for every regulatory system. How does this work? The GDPR mandates data deletion requests, while the CCPA requires explicit opt-out options, must be clearly defined procedures. This is particularly important.

*c) Regular Audits and Updates:*

Organizations need to do the audits and monitor compliance measures to remain up to date with regulatory changes. Both unified and jurisdiction-specific policies are reviewed as a part of this.

By covering all angles, compliance with both GDPR and CCPA can be figured out by introducing all-encompassing proposals that combine both jurisdiction-specific and generally established but often overlapping practices. The approach is designed to reduce the risk of legal and financial issues and, at the same time, to ensure the same level of data protection across different legal jurisdictions.

*C. Challenges and Solutions:*

The intense complexity and ambiguity of official terms make legal documents nearly impossible to completely and satisfactorily interpret through natural language processing (NLP). Many of the technical and highly complicated language in legal texts make it harder for machine learning models to understand and classify them appropriately. NLP models get stuck when they try to understand the context-sensitive nature of legal terms and phrases. A term's interpretation could change depending on its regulation text context.

Many options are available to handle these problems. These include:
1. Improving Model Training: One way to focus on the area of NLP models is to introduce new datasets that refer to the domain; this, in turn, will result in better accuracy. The model is getting a better feel for how legal terms and phrases are used in context during annotations made on the data sets. Through specialized training, the model's ability to understand and classify regulatory provisions is improved.
2. Incorporating Expert Feedback: The correct interpretation and dependability of NLP models could be significantly improved by using hand-refined text rather than the convoluted and choppy format provided by the NLP models. After checking the model's result, legal experts can rectify it and enhance its operation. The model is a very accurate representation of the legal language and context thanks to the iterative process.
3. Leveraging Ensemble Learning: Methods such as ensemble learning, which combine the numerous predictions of models to obtain greater precision and reliability, are gaining ground in machine learning. The fewer the confines of individual models, the more robustness the method supports. Through the utilization of ensemble learning, the model's capability to understand difficult legal texts is boosted, and this occurs due to the various models' strengths.
4. Implementing Explainable AI: Through the application of explainable AI techniques, the decision-making processes of machine learning models are explained. Hence, the absence of bias in regulatory analysis can be prevented by ensuring accountability through transparency. Using Explainable AI officers of compliance can gain insight into the model's predictions and improve confidence in the results. This has been the most important context in this regard.

5. Continuous Monitoring and Improvement: Legal frameworks keep changing as they are due to revisions and guidance being subject to them. NLP models must be continuously updated in order to stay accurate and relevant. Therefore, periodical audits and model reopening as the regulatory changes and the analysis still stay effective. The last sentence was edited for clarity.

It's important to note that NLP models can also be used as additional tools to support human experts. Large volumes of regulatory texts can be processed and analyzed by these models, but legal language is highly complex and context-specific, incessantly handled by humans.

1. Supplementary Assistance: NLP models are one of the digital solutions that can help regulatory compliance sectors recognize and segregate individual regulatory provisions, which leads to keeping the analysis part of the regulations as simple as possible. This way, human intervention is minimized, and areas that need to be reviewed by humans are identified. However, the real work is carried out by human professionals who should confirm and interpret the results for accuracy and harmony.
2. Human Oversight: The level of cognitive and contextual awareness that humans have is difficult to achieve with NLP models. They have developed the ability to cope with vagueness in legal language, make verdicts on the law texts, and produce a comprehensive analysis that considers the greater legal system among others. A combination of both NLP technology and professionals' skills leads regulatory compliance to be more efficient.
3. Continuous Collaboration: Maintaining Correctness, Consistency, and context-sensitivity of the solutions is impacted by the collaborative involvement of human experts in NLP models. Periodic remarks, insights, and advice from human professionals can boost the functionality as well as the dependability of the tools.

One of the best ways to become a regulatory advisor is to discuss the cooperation of NLP and human experts in the future. This suggests the value of joining technology with human checks to achieve the accuracy and efficiency of regulatory compliance.

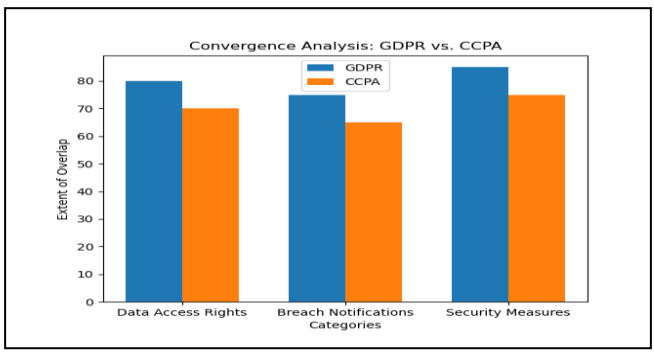

*Figure 3Convergence Analysis: Extent of Overlap in Data Access Rights, Breach Notifications, and Security Measures between GDPR and CCPA*

*D. Ethical and Privacy Considerations:*

The application of machine learning in regulatory compliance raises ethical and privacy concerns. Trust and privacy regulations can be maintained by making sure that the data used for model training is both anonymous and secure. Companies must introduce strong data governance measures to handle personal data effectively to meet both legal and ethical requirements. In addition, the models learned using machine learning need to be able to explain their decision-making process of the autonomous decision it has made to make them accountable and bias-free. Compliance officers and stakeholders involved in compliance processes can benefit from understandable AI analysis by further explanation of exact replicable methodologies and term definitions.

*E. Future Research Directions:*

The tenacity of AI for artificially creating compliance rules is not a sterile subject as of today; rather, there are a lot of new areas to study. To set the use of commonly accepted best practices in machine learning in the context of compliance analysis is a rule that must be obeyed so as to have uniformity and reliability. Processes and correctness, however, cannot be ensured if organizations do not create the required protocols for model construction, validation and deployment. In the case of another promising technology, such as blockchain, to ensure constant and visible record keeping throughout the compliance process will bring to the table overflow benefits. The combination of the use of blockchain and AI also offers a secure way of data storage while complying with regulations. In addition, there is a notable demand for the installation of NLP in legal texts. The efficiency of regulatory analysis can be heightened via the investigation of new methods of enhancing the accuracy and credibility (in domain-specific embeddings and context understanding techniques) of NLP models. Therefore, the NLP models by means of domain-specific embeddings and advanced contextual understanding techniques may be built such that their accuracy and credibility are enhanced in a way that could make regulatory analysis more efficient.

*F. Adaptive Compliance Framework for Future Research Directions:*

Multinational organizations have to deal with various issues where the ever-changing nature of regulatory frameworks is the most prominent one. For instance, compliance strategies that are static are becoming old school, since with the arrival of new changes, interpretations, and additional regulations by the entities the hidden defects of initial strategies are revealed. The suggested methodology is to integrate automatable regulatory updates with the machine learning models to have real-time analysis that can dynamically change. This strategy combines legal control functions and processes and encompasses the introduction of relevant ecological and economic factors from the legal operating aspect as well as the IT part, which ultimately leads to each company receiving relevant and rational instruments to act on.

Core Components:

*a) Regulatory Change Tracking System:*

The first part of the framework is to create a real-time regulatory monitoring system. This system keeps on scanning the official repositories EDPB and the Californian Legislative Databases to check the updates, new law, guidelines, as well as enforcement actions. The relevant changes are then logged and are going to be incorporated into the model update process.

*b) Dynamic NLP Model Update Mechanism:*

The framework incorporates a mechanism for dynamically retraining the NLP models whenever significant regulatory changes are detected. This involves automated pipelines for data preprocessing, annotation, and model fine-tuning. A robust validation process ensures that the retrained models maintain accuracy and reliability before deployment.

*c) Hybrid Human-AI Oversight Loop:*

Automation is, without question, a major part of the adaptive framework whereas human control can be called a driving force in case of large interpretations of regulations. Legal professionals make an occasional sneak peek at the most important developments and the models' production of the law, which in the meantime are made better by incorporating the received feedback. Under both forms of integrated approach, model precision is coupled with the experiential knowledge of the lawyers.

*d) Interactive Dashboard for Compliance Officers:*

To execute the policy, I suggest an interactive dashboard. This tool will show you the latest changes in law in real-time, will reveal similarities and differences in terms of compliance, and will provide legal compliance suggestions. Advisories regarding the most important updates in the legal field and the alterations of risk levels should help institutions streamline their tasks.

1. Workflow of the Framework:

Change Detection: The system keeps track of and documents the regulatory updates as they come in.

Model Retraining: The NLP model(s) are trained using the most recent data sets when significant changes are distinguished.

Expert Feedback: Only the most crucial information, which is given the expert review, is then modeled for the user to interface with the new recommendations.

Recommendation Generation: The up-to-date machine learning algorithms provide real-time slander and give some recommendations on actions.

Visualization: Compliance officers use the interactive dashboard to generate insights in the form of visualizations and alerts.

2. Benefits of the Adaptive Compliance Framework:

Real-Time Compliance Management: The system will inform the user about the changes in the regulations immediately, thus, legislation will be put into practice without any delay.

Cost Efficiency: The cost of compliance is lowered using Automation which slows down manual labor needed to track and interpret legal changes

Improved Compliance Accuracy: The nonstop model is run through training and expert feedback loops are put in place to make sure that a precise insight is provided.

Scalability: The structure has the potential to become bigger in size and through this organization it could be increased to include other regulations, which are particularly convenient for large multinational corporations

3. Challenges and Mitigation Strategies:

Data Availability: It is possible only on some jurisdictions having real-time legal data feeds. Partnering with legal data providers and creating direct integrations with official regulatory bodies are possible remedies.

Model Drift: Regular checking of the performance of the model by the audits which are scheduled and retraining of the model will help in the timely detection of the model's performance deviation due to the inclusion of outdated data or the shifting of the legal interpretation.

Reliance on Experts: Yet a mere reliance on experts is unworkable hence seeking cooperation with them is a must. As part of the time subsequently, feedback from the initial steps is vital to the identification of the model accuracy rate.

[12-14]

# VII. CONCLUSION

This research aims to utilize machine learning techniques, particularly natural language processing (NLP), to compare the GDPR and CCPA. Using sophisticated NLP models, this study has identified areas of convergence and divergence between the two major data privacy frameworks. The results highlight significant overlaps between both the GDPR and CCPA, particularly in relation to data access rights (including electronic receipts), data breach notifications, and data security measures. The shared requirements enable organizations to improve their compliance by implementing cohesive policies and procedures that cater to the common requirements of both regulations.

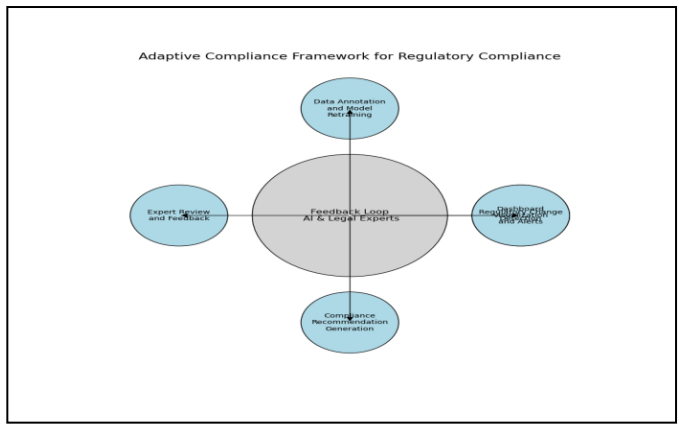

*Figure 4 Adaptive Compliance Framework for Regulatory Compliance*

Costs, operational efficiency, and data protection strategies can be boosted by this standardization. The analysis of divergences, however, shows that there are differences between each regulation, for instance the "right to be forgotten" GDPR regulation and an "opt-out of sale" provision in the CCPA. Specifically, these differing requirements require unique compliance strategies that cater to the specific needs of each regulatory framework. The CCPA mandates that organizations must provide consumers with explicit opt-out options regarding the sale of their personal information. In contrast, the GDPR requires organizations to facilitate data deletion requests under the 'right to be forgotten.' Moreover, it is essential to acknowledge the differences in scope, definitions, and enforcement mechanisms to ensure complete compliance with both regulations. The consequences of these findings are important for multinational companies operating in different countries. Several compliance strategies are developed to meet multiple legal obligations simultaneously, with the GDPR and CCPA providing guidance for organizations seeking to implement effective enforcement strategies. Nevertheless, specific techniques must be employed to cater to the particulars of each law, guaranteeing completeness. The implementation of NLP on legal texts poses several difficulties, such as the complexity and ambiguity of legal language. To overcome these problems, this study has proposed several solutions, including better model training with domain-specific datasets, the integration of expert feedback, ensemble learning methods, and the implementation of explainable AI. These techniques improve the precision and dependability of machine learning models used in regulatory analysis, and future research should focus on establishing standard practices for machine utilization in compliance analysis (including comprehensive guidelines for model design, validation, deployment, etc.). Furthermore, the integration of emerging technologies such as blockchain can be used to improve compliance processes by establishing permanent records of compliance actions. The use of machine learning-driven convergence analysis can lead to significant improvements in compliance with GDPR and CCPA. Through the integration of legal knowledge and technical approaches, this research facilitates the development of more efficient and effective compliance strategies, ultimately contributing to data privacy and protection.

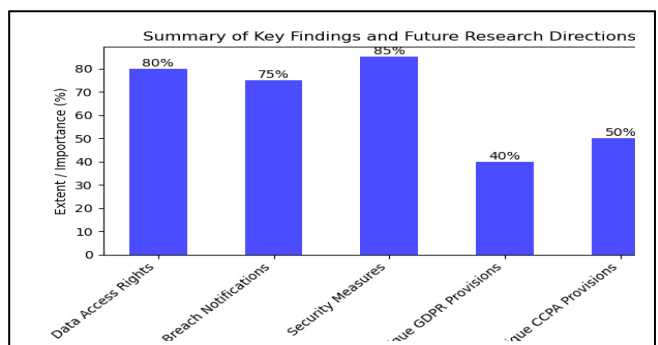

*Figure 5 Summary of Key Findings and Future Research Directions: Extent of Overlap and Unique Provisions of GDPR and CCPA*

[15-17]

## VIII. REFRENCES